# Superconducting micro-resonators for electron spin resonance - the good, the bad, and the future

Yaron Artzi,[1] Yakir Yishay,[1] Marco Fanciulli,[2] Moamen Jbara,[1] and Aharon Blank[1*]

1. Schulich Faculty of Chemistry

Technion – Israel Institute of Technology

Haifa 3200003

Israel

2. Department of Materials Science

University of Milano – Bicocca

Italy

*Corresponding author contact details: Aharon Blank, Schulich Faculty of Chemistry,

Technion – Israel Institute of Technology, Haifa 3200003, Israel,

phone: +972-4-829-3679, fax: +972-4-829-5948, e-mail: ab359@technion.ac.il.

https://orcid.org/0000-0003-4056-8103

# Abstract

The field of electron spin resonance (ESR) is in constant need of improving its capabilities. Among other things, this means having better resonators to reach improved spin sensitivity and enable larger microwave-power-to-microwave-magnetic-field conversion factors. Surface micro-resonators, made of small metallic patches on a dielectric substrate, provide very good absolute spin sensitivity and high conversion factors due to their very small mode volume. However, such resonators suffer from relatively low spin concentration sensitivity and a low-quality factor, a fact that offsets some of their significant potential advantages. The use of superconducting patches to replace the metallic layer seems a reasonable and straightforward solution to the quality factor issue, at least for measurements carried out at cryogenic temperatures. Nevertheless, superconducting materials, especially those that can operate at moderate cryogenic temperatures, are not easily incorporated into setups requiring high magnetic fields due to the electric current vortices generated in the latter's surface. This makes the transition from normal conducting materials to superconductors highly nontrivial. Here we present the design, fabrication, and testing results of surface micro-resonators made of yttrium barium copper oxide (YBCO), a superconducting material that operates also at high magnetic fields and makes it possible to pursue ESR at moderate cryogenic temperatures (up to ~80 K). We show that with a unique experimental setup, these resonators can be made to operate well even at high fields of ~1.2 T. Furthermore, we analyze the effect of current vortices on the ESR signal and the spins' coherence times. Finally, we provide a head-to-head comparison of YBCO vs copper resonators of the same dimensions, which clearly shows their pros and cons and directs us to future potential developments and improvements in this field.

# I. Introduction

The field of electron spin resonance (ESR) is in constant need of improving its capabilities. Among other things, this means having better resonators to reach improved spin sensitivity and enable larger conversion factors of microwave-power-to-microwave-magnetic-field (denoted as $C_p$). The sensitivity of electron spin resonance is determined, among other factors, by the quality factor, $Q$, of the resonator used in the experiment. From a quantitative standpoint, the sensitivity per unit of time is improved as $\sqrt{Q}$ [1, 2]. On the other hand, spin sensitivity also improves as the volume of the resonator, $V_c$, is reduced, with dependence described by $1/\sqrt{V_c}$ [3]. Unfortunately, the laws of nature determine that (generally speaking) the smaller the resonator's volume, the smaller its quality factor. Therefore, in many cases, when employing resonators with small $V_c$, some of the gains attributed to the reduction in size are offset by the loss in $Q$ (although $Q$ does not scale linearly with $V_c$, so the bottom line still shows some gains) [4-7]. This fact has motivated works on miniature resonators that would manage to maintain a high $Q$ [7]. One obvious way to improve the quality factor is to develop resonators made of superconducting materials [8-13]. Most relevant to this paper are works on surface micro-resonators that employ superconducting patches deposited on a dielectric substrate. For example, Lyon's group has employed a very thin and long ($\lambda/4$) microstrip resonator, made of Nb spattered on sapphire at a frequency of ~10 GHz, Q factor of ~1000–2000, temperature of ~4.2 K, and magnetic field of ~0.33 T to measure electron spins in P-doped isotopically enriched $^{28}$Si [9]. This was later extended to even smaller resonators with a sensitivity of ~$10^4$ spins and Q factor of 3100 measured at ~5 GHz (~180 mT) and 20 mK [14]. The group led by Cory has carried out similar work to that of [9], also with thin Nb microstrip resonators at similar frequencies and magnetic fields to show strong collective coupling to a sample of P-doped Si, with a Q factor of ~1500 [15]. Bertet's team has used micro-resonators made of aluminum deposited on silicon at a frequency of ~7.2 GHz, with Q factors of $7.8\times10^4$ up to $3\times10^5$ (with the lower Q value corresponding to the smaller resonator having better spin sensitivity), temperature of ~10 mK, and magnetic fields of a few millitesla, to measure a few tens of spins of bismuth dopants [16, 17].

The above-mentioned literature sources clearly show that while superconducting resonators have definitely been used in ESR in the past, they have mainly been employed in fundamental physics studies or proof of principle processes and are yet to penetrate more practical applications in mainstream ESR. This is because most studies have employed superconductors with very low critical temperatures, and mainly because unfortunately, superconducting materials and magnetic fields do not mix well. When a superconducting material is placed in a static magnetic field, it tends to lose its useful properties. Type I superconductors lose their superconductivity abruptly at rather low fields of a few millitesla. Type II superconductors are more forgiving, and materials such as yttrium barium copper oxide (YBCO) can maintain their superconductivity even at a field of ~150 T [18]. This property has already made YBCO and similar materials very useful for NMR and MRI applications, where the superconducting material is used as basis for the detection coil at a frequency of a few hundreds of megahertz [19]. YBCO has also the important advantage of acting as a superconductor even at moderate cryogenic temperatures (up to ~85K). Nevertheless, while the upper critical field (called $H_{c2}$) may be very large, there is another, lower, critical field, denoted $H_{c1}$, above which the superconductor is in a mixed state, with larger losses and small persistent current vortices (called Abrikosov vortices). Typical values of $H_{c1}$ for YBCO range at ~5–15 mT [20]. Thus, for a typical configuration where a thin (~100 nm) superconducting material layer is deposited on a flat substrate, the magnetic field *perpendicular* to the surface should not be larger than $H_{c1}$ or else losses will increase, and current vortices will cause constant fluctuation in the magnetic fields experienced by the sample (if it is placed near the surface of the superconducting layer). For NMR and MRI, which operate at low frequencies with relatively large coils, the increased loss issue is less of a problem, and samples can be placed at some distance from the coil, so that current vortices do not affect them. For ESR, however, this is a bigger issue, both due to increased losses at high microwave frequencies and mainly because of the necessity to place the sample very close to the superconducting surface to avoid an excessive drop in sensitivity.

In view of these issues, most of the previous studies on ESR with superconducting resonators, such as those cited above, are limited to very low static magnetic fields and extremely low temperatures [16], employ resonator structures that are thin and long[1] to avoid magnetic field effects as much as possible, or require a careful and painstaking alignment process (cooling → checking Q values → warming above $T_c$ → re-aligning and vice versa, until no vortices are generated) to place the direction of the static magnetic field exactly in the plane of the superconducting surface of the resonator [9, 21]. On the other hand, as noted above, the field of NMR has long ago showed that superconducting structures can contribute immensely to measurement sensitivity in mainstream applications [22], even at fields of more than 10 T. Clearly, there is still much more work to be carried out to unleash the full benefits of superconductivity upon the field of ESR.

*Motivation:* Our overall motivation is to develop optimized resonators for *mainstream pulsed ESR applications*. By "optimized" we refer to resonators having excellent spin and concentration sensitivity that can also provide a high conversion factor, $C_p$, from microwave input power to microwave magnetic field ($B_1$), to obtain an efficient pulse operation. The term

---

[1] Thin and long resonators are good for mitigating the effects of static magnetic fields. However, they are far from optimal for measuring "normal" samples that often have the same dimensions in all axes rather than being thin and long themselves.

"mainstream ESR" is of course not well defined, but our subjective definition relates to static fields in the range of 0.3–3.5 T and temperatures in the range of ~4–80 K, where one could probably find more than 80% of contemporary pulsed ESR works used in biology, chemistry, and materials science. Clearly, most, if not all, of the above-mentioned experiments with superconducting resonators in ESR are irrelevant to mainstream pulsed ESR. This is either because of the required extremely low temperatures, the devices' limitation of magnetic fields to the lower ranges, the resonators' Q values that are too large to excite large bandwidths, problems of nonlinearity while handling high microwave power, and/or because of their very poor concentration sensitivity. We argue that surface micro-resonators designed for the measurements of sample volumes of ~1 nL are possibly the best candidates for such mainstream pulsed ESR experiments [3, 6, 23]. This is because they exhibit both excellent spin and concentration sensitivity (~$5\times10^6$ spins/√Hz and ~0.03 μM/√Hz, respectively), as well as very high $C_p$ (at least ~10 G/√W, and often much more, as we show below for the present work). This makes pulsed ESR operation very efficient, mainly because it is possible to use tiny samples with very low input power (~10–100 mW), which enables the simple use of low-saturation-power cryogenic preamplifiers without the need for protections and limiters, thus providing a further boost in sensitivity. The reason we relate to a sample volume of ~1 nL is because much smaller volumes would result in (too) poor concentration sensitivity for most applications, while much larger volumes would result in too low spin sensitivity and much reduced $C_p$ that would require a much higher input power and thus would limit the advantages of cryogenic preamplifiers [3, 24]. Therefore, within this scope of surface micro-resonators, our specific aim here is to find ways to further improve their spin and concentration sensitivities, as well as increase their $C_p$. However, we want to do all that under the condition of being able to work at the full range of static magnetic fields and temperatures we specified above ***for*** *mainstream pulsed ESR*. For this reason, we resort to the use of superconducting materials, and specifically YBCO, which complies with our definition of mainstream pulsed ESR operating conditions. The existing literature on ESR with YBCO resonators is very scarce, with researchers using mostly simple half-wavelength microstrip resonators (see, for example, [25]) without providing information about vortices-induced magnetic field inhomogeneities and their effects on the sample's $T_2$. Such information (especially regarding $T_2$ changes in the presence of superconductors) is not found in the literature for any type of superconducting material. Moreover, we are not aware of attempts to produce micro-resonators made of YBCO.

Here, we present our efforts to answer the challenges brought forward in the Motivation section above. We show the design, production, and testing of a superconducting version of the "ParPar" family of surface micro-resonators recently introduced by us [23], operating at Q-band (magnetic field of ~1.25T corresponding to ~35 GHz). The superconducting resonators described here are based on YBCO-on-LaAlO$_3$ single crystal structures to facilitate measurements at high magnetic fields and at relatively high temperatures. We first outline the design process carried out by our own homemade finite-element software, which takes into consideration the unique electromagnetic properties of superconducting materials. Following that, we describe the fabrication procedure for the resonators. The produced resonators are then tested for their electromagnetic properties (resonance frequency and quality factor, at various static magnetic fields), as well as for their performance in detecting ESR signals. Issues caused by the static magnetic field and the latter's effects on ESR signal linewidth and $T_2$ are presented and experimentally characterized, followed by a discussion and a demonstration of methods to mitigate them. Finally, we compare the sensitivity and the $C_p$ of one of the superconducting micro-resonator structures to that of a similar resonator, made from copper, and draw general conclusions.

## II. Materials and methods

i. Test sample: All ESR measurements were carried out with a test sample of phosphorus-doped isotopically-enriched $^{28}$Si (denoted as $^{28}$Si:P). The P:$^{28}$Si epilayer is 50 μm thick and was grown using $^{28}$SiH$_4$ on a Si(100) *p*-type highly resistive substrate made by ISONICS Corporation (USA). The concentration of $^{29}$Si in the P:$^{28}$Si epilayer is below 0.1%. The phosphorus concentration was measured using the Hall effect and found to be 3.3(4)×10$^{16}$ cm$^{-3}$ [26].

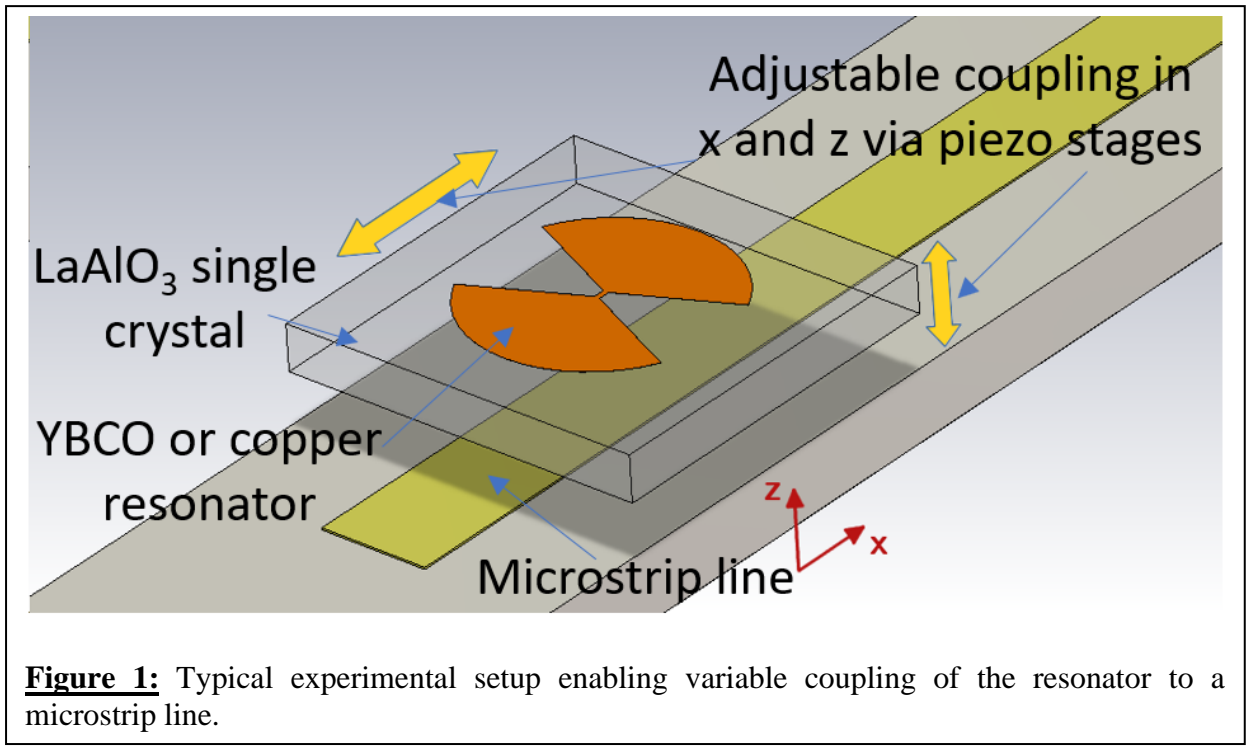

**Figure 1:** Typical experimental setup enabling variable coupling of the resonator to a microstrip line.

ii. ESR system and cryogenic probe head: Test sample measurements were carried out using a spinUP system equipped with spinUP-Q Q-band transconverter module (spinflex, Israel), with a Varian E-12 magnet and E-7700 power supply. The cryogenic measurements employed a homemade cryogenic probe head for Q-band (see Fig. 1 in ref [27]). It included a cold 10-dB attenuator on the input power line to eliminate external room temperature noise. Excitation power was measured using a diode detector and a 10-dB directional coupler to match the nominal power input to the resonator; this measurement took into account the 10-dB cold attenuator, but not additional line losses from the bridge to the resonator (measured at ~8 dB). In our present experiments, the imaging coil module in the said probe head was replaced by a cylindrical copper shield with inner diameter of 3 mm and outer diameter of 9.5 mm. Moreover, inside the cryostat we placed a small Hall sensor (model HE144P made by Advanced Sensor Technology, Sweden), which—unlike silicon-based chips—is able to operate at cryogenic temperatures. This Hall sensor measured the out-of-plane static magnetic field component experienced by the resonator and was placed ~8 mm away from it. The resonator was excited by a microstrip line with variable coupling controlled by attocube systems' piezo stages (See Fig. 1 here and Fig. 1 in ref [27]). Cooling to 10 K was achieved using a Janis cryostat (model STVP-200). An additional important component of our setup was a coil placed just outside the cryostat whose axis was coaxial to the cryostat and perpendicular to the direction of the static magnetic field. This coil can sustain currents of up to ±5 A and produce a magnetic field of ~70 G per A with the purpose of canceling any out-of-plane static magnetic field component experienced by the resonator.

## III. Resonator design

The resonators are designed by means of a custom-made integral equation electromagnetic software we recently developed in our lab. Full details of this program are provided in [28]. Briefly, it employs a method-of-moments (MoM) solver based on the electric field integral equation. The software can account for either conducting or superconducting layers deposited on a dielectric substrate. Conductor or superconductor materials are modeled through their surface impedance, which is given by eqs. (1) and (2) for conductors and superconductors, respectively [29].

(1)
$$Z_s = \frac{\kappa}{\sigma} \frac{e^{\kappa t_s} + \frac{\sigma\eta - \kappa}{\sigma\eta + \kappa} e^{-\kappa t_s}}{e^{\kappa t_s} - \frac{\sigma\eta - \kappa}{\sigma\eta + \kappa} e^{-\kappa t_s}},$$

(2)
$$Z_s = j\omega\mu_0\lambda_L \frac{e^{t_s/\lambda_L} + \frac{\eta - j\omega\mu_0\lambda_L}{\eta - j\omega\mu_0\lambda_L} e^{t_s/\lambda_L}}{e^{t_s/\lambda_L} - \frac{\eta - j\omega\mu_0\lambda_L}{\eta - j\omega\mu_0\lambda_L} e^{t_s/\lambda_L}},$$

where $\eta$ is the dielectric medium impedance, $\sigma$ is the complex conductivity, $\kappa = (1+j)\sqrt{(\omega\mu_0\sigma)}/\sqrt{2}$, and $\lambda_L$ is the London penetration depth of the superconductor. Here we used $\lambda_L$ = 500 nm, which gave good correspondence between the calculated and measured results of the resonance frequencies. (Literature data on $\lambda_L$ vary a lot [30, 31], but some sources provide values similar to those found here [32]) The representation of the conductors/superconductors by their surface impedances is crucial, as it makes it possible to overcome numerical difficulties associated with solving resonators comprising extremely thin layers and dielectric substrates. In practice, we were able to produce accurate electromagnetic field solutions which could not be obtained by modeling the conductors/superconductors as closed objects having finite thickness.

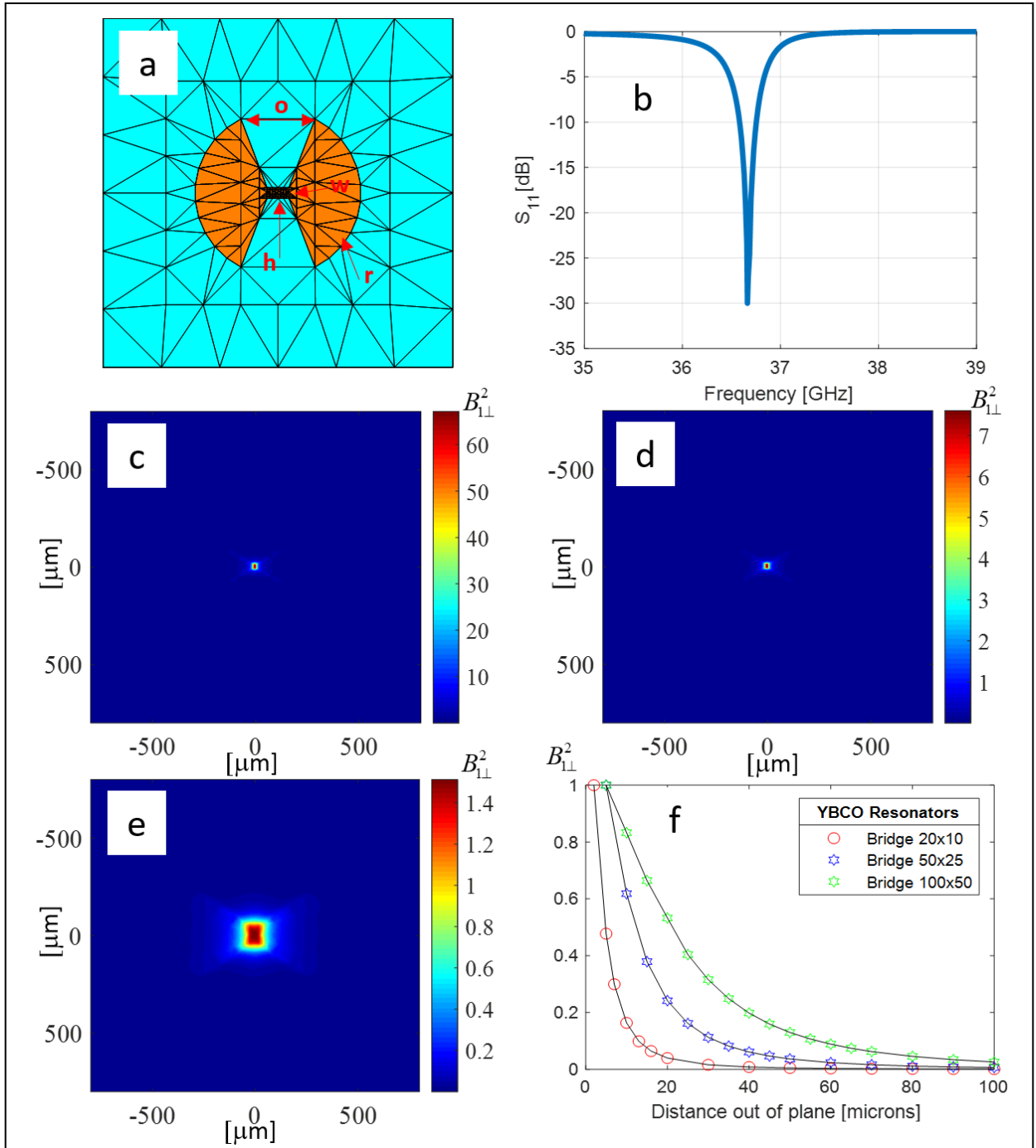

**Figure 2:** (a) Geometry of the ParPar resonators, as used in our electromagnetic solver, showing also the resonator's geometry parameters described in Table 1. (b) Calculated reflection coefficient ($S_{11}$) parameter for the ParParYBCO-20 resonator. (c) Calculated magnetic field energy for fields perpendicular to $B_0$ (denoted as $B_{1\perp}^2$) for the ParParYBCO-20 resonator, at a height of 5 μm above the resonator's surface. (d) The same as (c), but for the ParParCopper-20 resonator. (e) The same as (c), but for the ParParYBCO-100 resonator, at a height of 25 μm above surface. The color bar in (c), (d), and (e) shows the increase in $B_{1\perp}^2$ for a smaller resonator and for a larger Q. (f) Calculated $B_{1\perp}^2$ at the center of the resonator, as a function of height above the resonator's surface, for 3 different types of YBCO resonators.

The electromagnetic solver we developed was employed in the design of several structures, as depicted in Fig. 2 and summarized in Table 1. While ParPar resonators can be fabricated in a range of bridge dimensions, from ~1 μm to ~200 μm, we chose here to work with bridge dimensions of only a few tens of microns. This is because, as discussed above, very small bridge dimensions result in excellent absolute spin sensitivity but poor concentration sensitivity [3, 33, 34], which is not suitable for most general-purpose ESR applications. Nevertheless, as we shall see below, very small bridge dimensions may resolve some of the difficulties associated with the operation of superconducting resonators under high static magnetic fields. All calculations were carried out with plane wave excitation, which generates a slightly larger resonance frequency compared to the excitation by microstrip line actually measured in our work.

| Resonator's Name | Conducting Material | r [$\mu m$] | o [$\mu m$] | h [$\mu m$] | w [$\mu m$] | Q | T [K] | Res. freq. (calc.) [GHz] | Res. freq. (meas.) [GHz] |
|---|---|---|---|---|---|---|---|---|---|
| ParParCopper-20 | Copper | 380 | 414 | 20 | 10 | 30/130 | 298/10 | 38.1 | 36.3 |
| ParParCopper-50 | Copper | 380 | 417 | 50 | 25 | 30/150 | 298/10 | 37.7 | 36.4 |
| ParParYBCO-20 | YBCO | 365 | 385 | 20 | 10 | 100/1030 | 77/10 | 36.7 | 35.4 |
| ParParYBCO-50 | YBCO | 360 | 427 | 50 | 25 | 100/1030 | 77/10 | 37.1 | 35.5 |
| ParParYBCO-100 | YBCO | 370 | 433 | 100 | 50 | 200/1390 | 77/10 | 36.8 | 35.7 |

**Table 1:** Resonator properties relevant to this work (refer to Fig. 2a for the dimension parameters). The measured Q values and resonance frequencies are typical values since they are coupling-dependent. The calculated resonance frequency is based on a plane wave excitation model.

## IV. Resonator fabrication

The ParParCopper resonators were produced using the process described in ref [33] but with a 200 μm- thick LaAlO$_3$ substrate (purchased from MTI, USA). The ParParYBCO resonators were built using a 200 μm-thick LaAlO$_3$ substrate covered with a 40-nm CeO$_2$ matching layer, 100-nm YBCO layer, and a 100-nm CeO$_2$ capping layer (samples purchased from Ceraco, Germany). LaAlO$_3$ was chosen as a substrate because of its fairly large permittivity that is almost temperature-independent and its low dielectric losses ($\varepsilon \sim 24$, tan $\delta < 10^{-4}$) [35]. The larger the substrate's permittivity, the more condensed the mode volume of the resonator, which makes it possible to obtain a relatively large microwave magnetic field, $B_1$, just above the resonator's bridge region. Moreover, its lattice constant is similar to that of YBCO, which allows for stable growth of the superconducting layer. The YBCO thickness of 100 nm was chosen after performing preliminary experiments with thickness of 50, 100, and 200 nm. It was found that at a thickness of 50 nm, Q was too low and showed less sensitivity to magnetic field effects, while at a thickness of 200 nm, Q was higher but more sensitive to those effects. Therefore, a thickness of 100 nm was considered to be a good compromise. We fabricated the resonators with substrate dimensions of $1.6 \times 1.6 \times 0.2$ mm, various bridge dimensions, and various diameters. To etch the samples into shape, we deposited a thick layer of photoresist on them (~2.5 $\mu$m) for protection, and patterned it in a standard microphotolithography process (AZ1518 photoresist, spin-coat at 2000 rpm; pre-bake: 110 ºC, 2 min; exposure: 274 nm, 600 mJ/cm$^2$, 1.8 s). We then performed ion-milling with an AJA International Inc. ATC-IM miller (with the following parameters: argon ions; discharge voltage: 40 V; beam voltage: 500 V; beam charge: 90 V; acceleration voltage: 120 V) to etch away the unwanted YBCO. From calibration measurements we estimated that in these conditions we could etch CeO$_2$ at 0.8-1.2 Å/s and YBCO at 0.4−1 Å /s, so we ran ion-milling processes of 3000 s in 12 rounds of 250 s each (to avoid overheating the samples) and achieved the required YBCO patterning. Each wafer ($15 \times 15$ mm) contained an average of 49 resonators that were diced using a Disco DAD3350 dicer.

## V. Resonator tests

Following the fabrication of several types of resonators, individual devices were glued to a custom-made Rexolite resonator holder using a cryogenic adhesive (Stycast 1266) for use in our cryogenic ESR probe head [23, 33]. The resonators were tested for their electromagnetic and ESR-related properties at various temperatures and under different static magnetic fields.

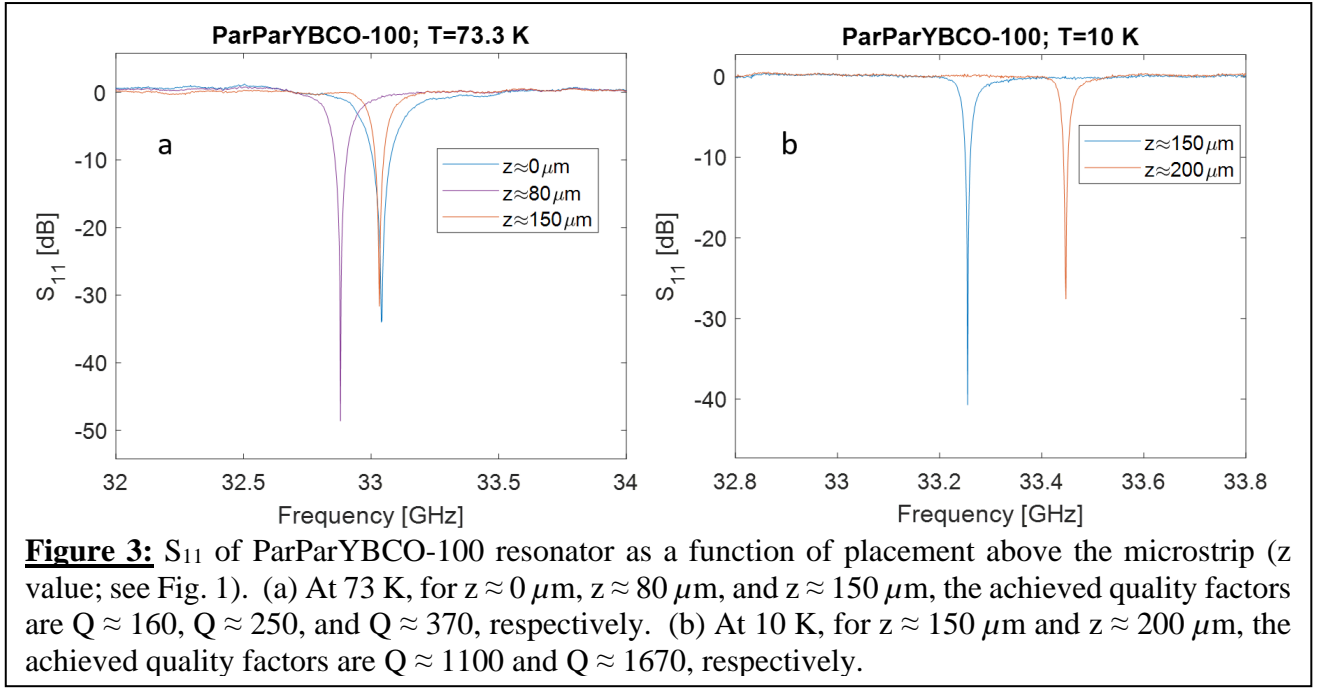

**Figure 3:** $S_{11}$ of ParParYBCO-100 resonator as a function of placement above the microstrip (z value; see Fig. 1). (a) At 73 K, for $z \approx 0$ $\mu$m, $z \approx 80$ $\mu$m, and $z \approx 150$ $\mu$m, the achieved quality factors are Q ≈ 160, Q ≈ 250, and Q ≈ 370, respectively. (b) At 10 K, for $z \approx 150$ $\mu$m and $z \approx 200$ $\mu$m, the achieved quality factors are Q ≈ 1100 and Q ≈ 1670, respectively.

a. Tests of electromagnetic properties

Fig. 3 shows typical results of the reflected power ($S_{11}$) measured (using Agilent ET8600 VNA) on one of our ParParYBCO-100 resonators at 77 K and 10 K. It is evident that at these temperatures, our setup can reach Q values of up to 1670. Moreover, the adjustable coupling with the piezo stages allows for relatively wide tunability of the resonance frequency, up to ~200 MHz, with relatively small effects on Q. Following these tests, we proceeded to examine the influence of the magnetic field on the resonator's properties. The resonator was placed in the probe head in such a manner that the static magnetic field from the electromagnet was parallel to the device's bridge. Thus, ideally, there should have been no magnetic field component in the direction perpendicular to the resonator's surface. This way, it was hoped, no current vortices would form in the resonator, and we would expect to see minimal effects of the magnetic field on the resonator's properties. Figure 4 shows that while the resonance frequency did not change dramatically as a function of the magnetic field, the quality factor did drop by about ~20%. This behavior implies that vortices do form in the resonator's superconducting layer, as is also evident from the ESR experiments performed by us (see next section). The slight increase in Q value at the beginning of the curve is attributed to coupling changes due to the effect of the static field on the circulator positioned inside the probe head, which is slightly affected by the static field.

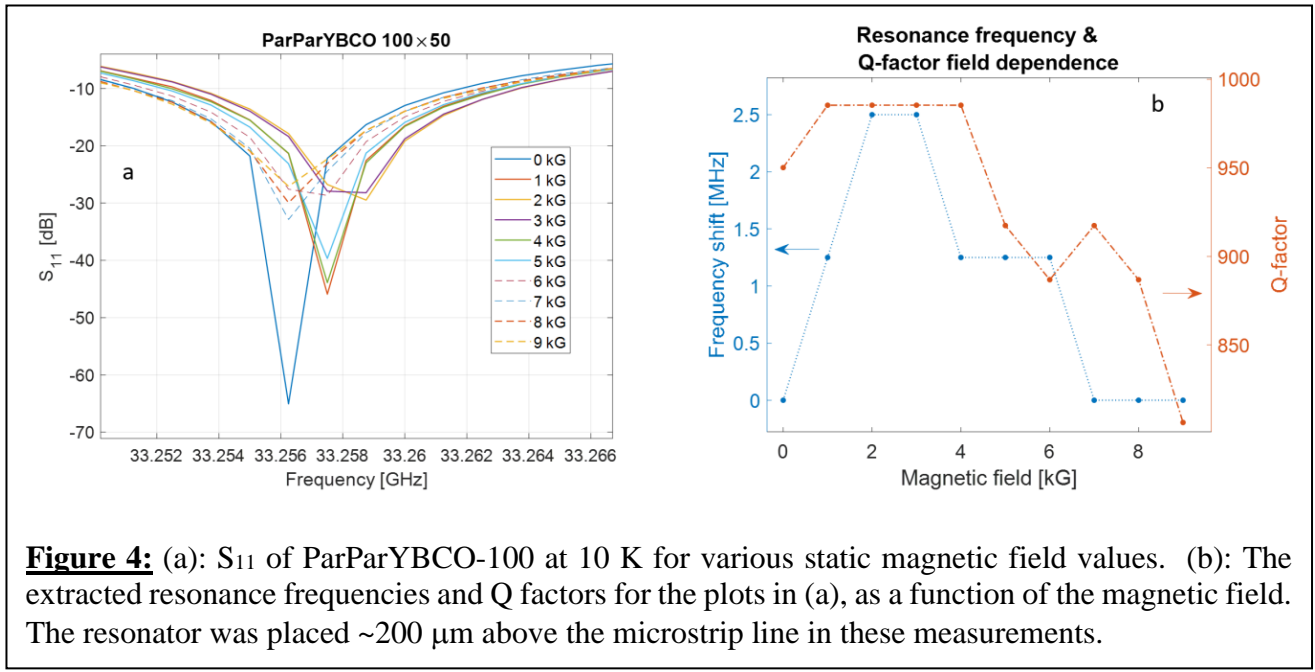

**Figure 4:** (a): S$_{11}$ of ParParYBCO-100 at 10 K for various static magnetic field values. (b): The extracted resonance frequencies and Q factors for the plots in (a), as a function of the magnetic field. The resonator was placed ~200 µm above the microstrip line in these measurements.

b. Tests of ESR properties

Following the measurements of the resonators' fundamental electromagnetic properties, further experiments were conducted to learn about their ESR performances. Our ESR tests aimed to (i) evaluate the effect of current vortices on the ESR signal's magnitude, spectral width, and on the sample's coherence properties; (ii) attempt to mitigate the effect of vortices by canceling the out-of-plane static magnetic field component; and (iii) compare the ESR signal and spin sensitivity of similar resonators made of normal conducting (copper) and superconducting (YBCO) materials.

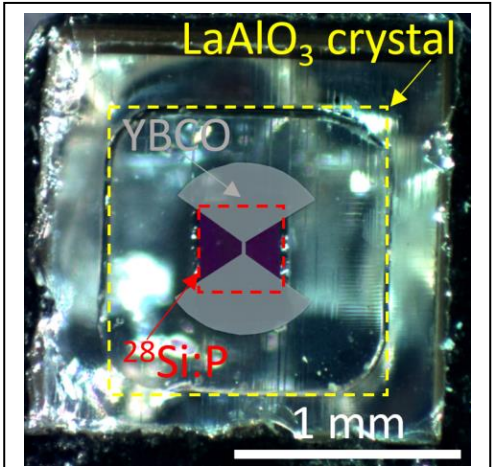

**Figure 5:** Bottom view of the ParParYBCO-50 resonator with the $^{28}$Si:P sample (seen through the resonator's crystal). For illustration purposes, the resonator and sample are marked with false colors.

*(i) Effect of current vortices on the ESR signal and on the sample's coherence properties:* Our hypothesis was that current vortices cause magnetic field distortion near the resonator's surface that can broaden the ESR signal and also affect its coherence properties. To test this hypothesis, we employed the $^{28}$Si:P sample described above, which has two very narrow lines in its spectrum and a long coherence time ($T_2$). Therefore, such sample would be very sensitive to magnetic field distortions. Fig. 5 shows the sample placed on one of our resonators (ParParYBCO-50). The sample dimensions were considerably larger than the bridge dimensions (in this case, $25 \times 50$ µm, see Fig. 2), so we could excite spins close to the bridge with low-power microwave excitation, and also peripheral spins (further away from the bridge and superconducting layer) with higher-power microwaves. This was verified in our past work with ESR microimaging experiments (see refs [23, 27]). To make a fair comparison, we compared the ESR spectra of normal conducting and super conducting surface micro-resonators. Figure 6 shows typical experimental results for the field-swept ESR spectrum of the $^{28}$Si:P sample for the ParParCopepr-50 and ParParYBCO-100 resonators. It is evident that the superconductor material greatly broadens the ESR signal. Moreover, as we use less power for pulse excitation (~0.1 mW vs ~1.1 mW), the signal broadens even more, probably due to the increasing effect of vortices close to the resonator's surface. The effect of current vortices is also clearly visible through the echo decay times, as depicted in Fig. 7. While for the copper resonator we see that $T_2$ slightly increases as we go down in power (probably due to smaller effect of instantaneous diffusion (see ref [36], page 216 , and also [37]) ), we clearly see a sharp decrease in $T_2$ for the superconducting material as power is reduced, especially with no current bias on the external coil perpendicular to $B_0$ (see also below). Another point of interest in Fig. 6 is the apparent increase in the central line's magnitude, relative to the two outer peaks, when measuring at lower power. This central line appears for pairs of closely spaced P atoms [38]. One possible way to interpret this increase is to assert that the concentration of the P atoms is inhomogeneous and is larger close to the surface. This can also explain the reduction in $T_2$ at low excitation power and may make our arguments regarding the effects of the current vortices redundant. However, such relative increase of the central peak was not seen in the copper resonators and thus we can conclude that the vortices' effects are real and that the dimers' line is possibly less sensitive to them and thus becomes relatively larger, compared to the main outer spectral peaks, at lower excitation powers.

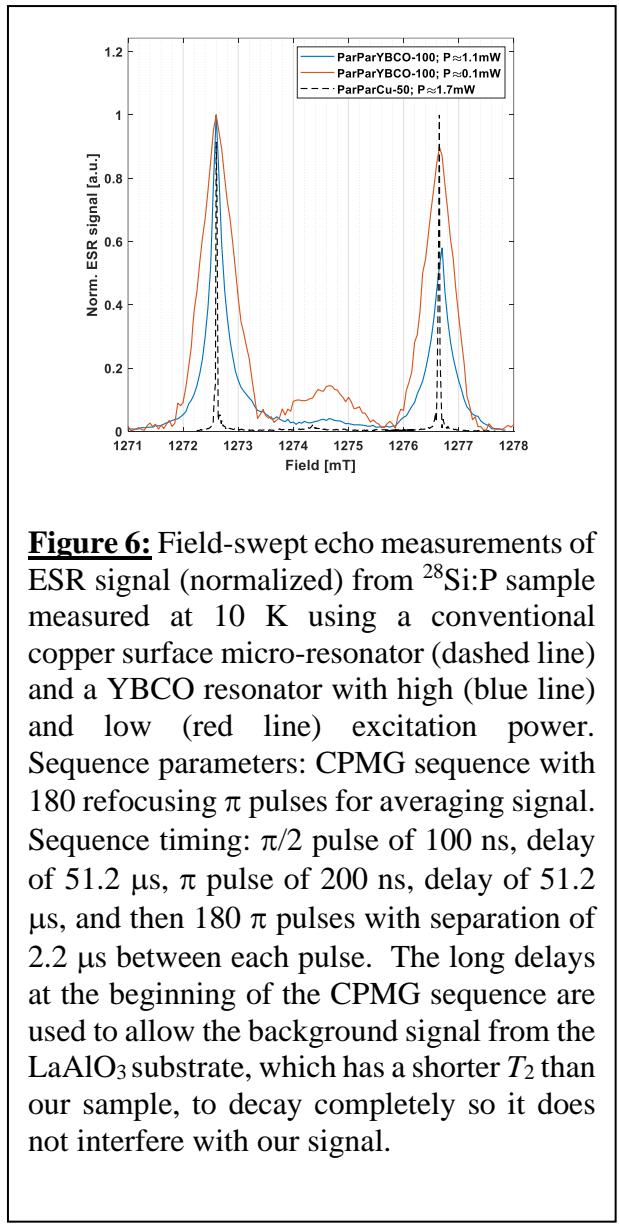

**Figure 6:** Field-swept echo measurements of ESR signal (normalized) from $^{28}$Si:P sample measured at 10 K using a conventional copper surface micro-resonator (dashed line) and a YBCO resonator with high (blue line) and low (red line) excitation power. Sequence parameters: CPMG sequence with 180 refocusing π pulses for averaging signal. Sequence timing: π/2 pulse of 100 ns, delay of 51.2 μs, π pulse of 200 ns, delay of 51.2 μs, and then 180 π pulses with separation of 2.2 μs between each pulse. The long delays at the beginning of the CPMG sequence are used to allow the background signal from the LaAlO$_3$ substrate, which has a shorter $T_2$ than our sample, to decay completely so it does not interfere with our signal.

*(ii)    Mitigating the effects of current vortices*

Having realized that current vortices greatly affect our sample's ESR signal, attempts were made to mitigate their effect. As noted above, the usual practice is to employ a resonator with a thin superconducting layer and align its plane along the direction of the static magnetic field. However, while previous works provided mainly mechanical solutions involving a repeated process of aligning → reducing the temperature below $T_c$ → measuring Q → elevating the temperature above $T_c$ to anneal out the vortices → realigning [9], and so on, we tried to find a different solution. We employed an in-probe Hall sensor and additionally had an external coil to try to cancel out all residual static field components perpendicular to the plane of the superconductor. Initially, we tested this procedure by setting $B_0$ to its nominal value → setting the field bias of the external coil to cancel out the reading in the internal Hall sensor → lowering the temperature below $T_c$ → testing the quality of the ESR signal (the magnitude and the width of the resonance line). Unfortunately, this process did not provide good enough results at first shot as we expected, probably due to the slightly different locations of the Hall sensor and the resonator. Nevertheless, surprisingly enough, we found out that it was possible to adjust the bias of the external coil in real time, without having to raise the temperature and anneal the vortices, and thus get an immediate feedback (via the ESR signal) to the quality of our out-of-plane field cancelation. Typical results of such tests conducted with the ParParYBCO-100 resonator at 10 K are provided in Fig. 8. It was clear that the bias greatly affected the existence of vortices, and that at an optimal bias the observed line width was almost as narrow as the one obtained using the copper resonators (see also below). Additionally, some hysteresis effects were observed, where the current bias that produced the narrowest line was found to be dependent on the current the system was biased to prior to setting that current. In general, increasing the current and then decreasing it resulted in a higher optimal bias current (i.e., that which produced the narrowest line) than when the current was first decreased and then increased. We also inspected the effect the bias current had on the resonator's $S_{11}$ parameter, but we did not observe any clear indication that at some bias field the Q factor reaches a maximum value. We did observe clear changes in resonance frequency and some slight changes in Q factor; however, these were not useful in determining the optimal bias. Possibly, the approach to Q-value monitoring that was used in the past for mechanical alignments is more useful to identify the optimal bias in higher Q-value resonators (achievable at lower temperatures). Nevertheless, as noted above, in our case the ESR signal itself served directly as our feedback input and we had no need to apply this Q-value-based aligning approach.

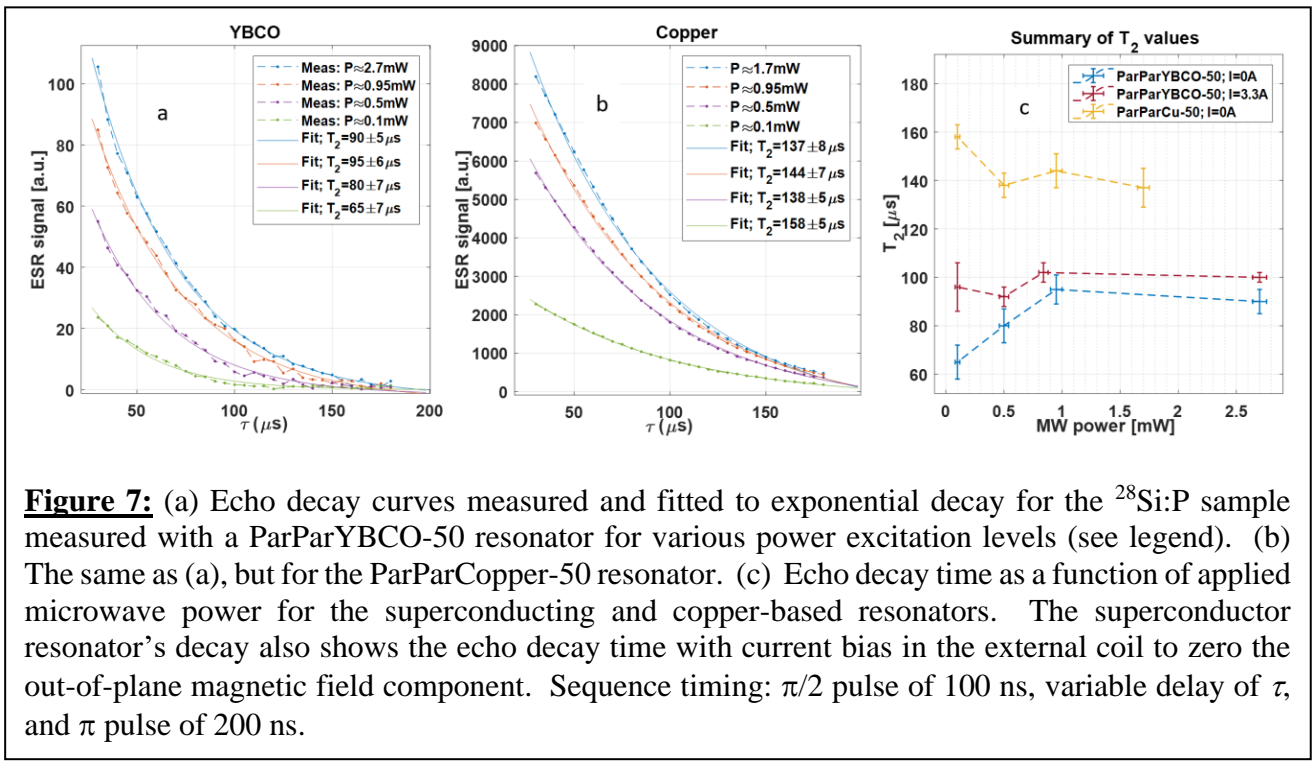

**Figure 7:** (a) Echo decay curves measured and fitted to exponential decay for the $^{28}$Si:P sample measured with a ParParYBCO-50 resonator for various power excitation levels (see legend). (b) The same as (a), but for the ParParCopper-50 resonator. (c) Echo decay time as a function of applied microwave power for the superconducting and copper-based resonators. The superconductor resonator's decay also shows the echo decay time with current bias in the external coil to zero the out-of-plane magnetic field component. Sequence timing: π/2 pulse of 100 ns, variable delay of τ, and π pulse of 200 ns.

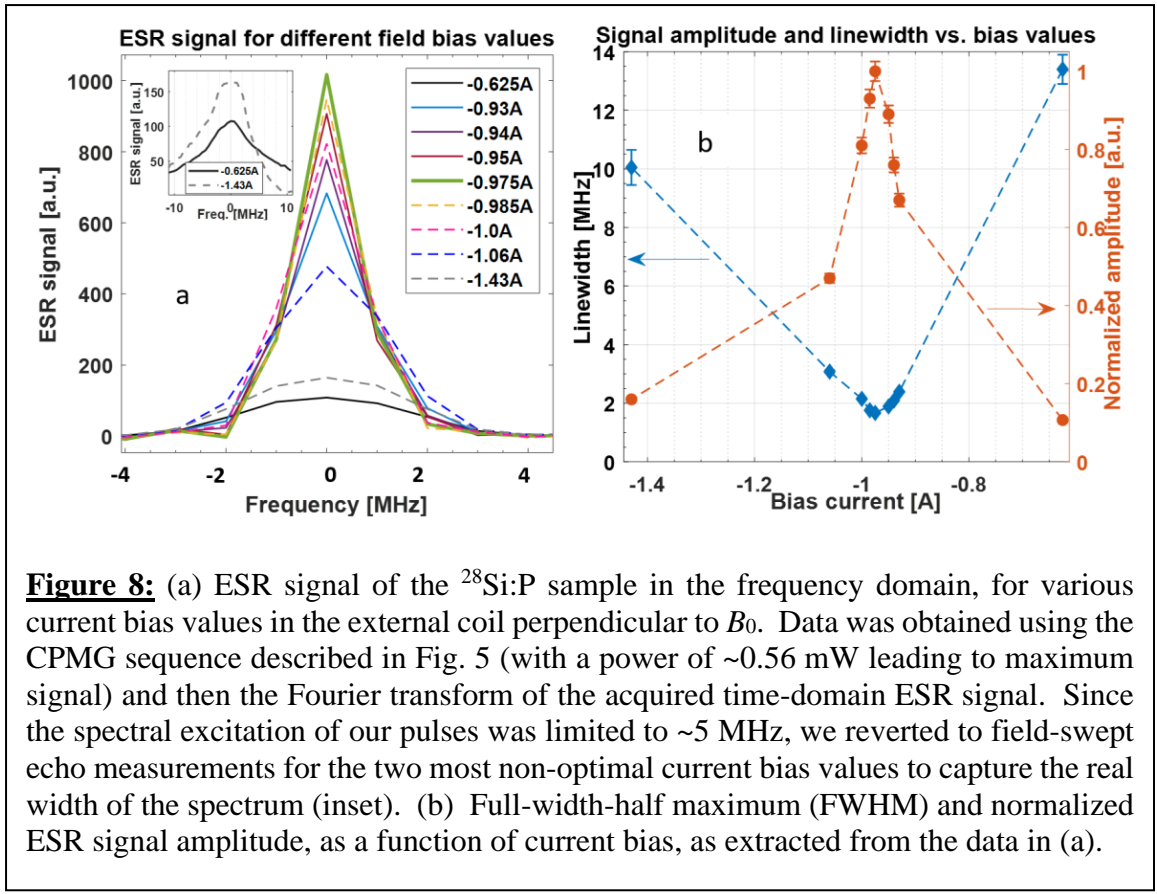

**Figure 8:** (a) ESR signal of the $^{28}$Si:P sample in the frequency domain, for various current bias values in the external coil perpendicular to $B_0$. Data was obtained using the CPMG sequence described in Fig. 5 (with a power of ~0.56 mW leading to maximum signal) and then the Fourier transform of the acquired time-domain ESR signal. Since the spectral excitation of our pulses was limited to ~5 MHz, we reverted to field-swept echo measurements for the two most non-optimal current bias values to capture the real width of the spectrum (inset). (b) Full-width-half maximum (FWHM) and normalized ESR signal amplitude, as a function of current bias, as extracted from the data in (a).

*(iii)   ESR signal and spin sensitivity of copper vs YBCO resonators*

After eliminating almost all of the effects of the current vortices, the main question remains: is it worth it? Is the use of YBCO surface resonators justified, at least for the conditions we tested here and for the type of sample we measured? To answer this question, we carried out a head-to-head comparison of ParParCopper-20 and ParParYBCO-20 resonators. Both have the same bridge size and similar resonance frequency but slightly different overall dimensions, due to the issue of kinetic inductance, which is taken into account by our electromagnetic solver. Both resonators were tested with the exact same sample of $^{28}$Si:P, as shown in Fig. 5, placed on their surfaces. Fig. 9 shows the measured frequency domain ESR signal of both resonators, along with the noise spectrum, acquired off-resonance. It is evident that the copper resonator has a larger signal peak and narrower spectrum; however, its noise level is slightly larger (by a factor of ~1.2) than that of the YBCO resonator. We have seen a similar trend of reduced noise in several measurements taken from the superconducting resonators. As for the SNR itself, to accommodate for the different line widths we can compare two cases, (i) SNR when signal and noise are both considered for a single frequency bin (where the signal is maximal), and (ii) SNR when signal and noise are both averaged over 3 frequency bins. In case (i) we get an SNR of 8830 and 6500 for the copper and YBCO resonators, respectively. In case (ii) the corresponding results are 6730 and 6820. It probably would have been better to compare also the case of noise averaged over only 2 frequency bins; unfortunately, we do not have a recording of the signal that would work well for this case (where the signal is symmetrically distributed between two nearby frequency bins).

In terms of absolute spin sensitivity, we can provide the following estimations: For the copper resonator, the maximal signal was obtained with an input microwave power of ~3.6 mW and, as noted above, it gave an optimal SNR of 8330. Considering our microwave magnetic field simulations and past magnetic resonance imaging performed with resonators with the same dimensions [23], we can estimate that the excited volume in this case was roughly ~70 × 70 × 35 μm$^3$ (the microwave magnetic field, $B_1$, is very inhomogeneous and degrades rapidly with height above the resonator's surface). With our $^{28}$Si:P sample, the above volume corresponds to ~5.7 × 10$^9$ spins, which produce the abovementioned SNR in one second of averaging. This means that the absolute spin sensitivity of this resonator is ~6.9 × 10$^5$ spins/√Hz. A similar analysis provides an absolute spin sensitivity of ~8.2 × 10$^5$ spins/√Hz for the YBCO resonator. In this calculation we assume that the volume of the sample excited by the YBCO resonator was the same as the volume excited by the copper resonator, although the required excitation MW power was only ~1.07 mW. This is understandable since in our measurements the ESR signal grew with power and saturated at such a low level for the YBCO resonator; thus, we assume that because of the high Q, we need a much smaller power (the conversion factor goes as the square root of Q). In such case, the excited volume should be similar to the one achieved in the copper resonator with an excitation power of 3.6 mW.

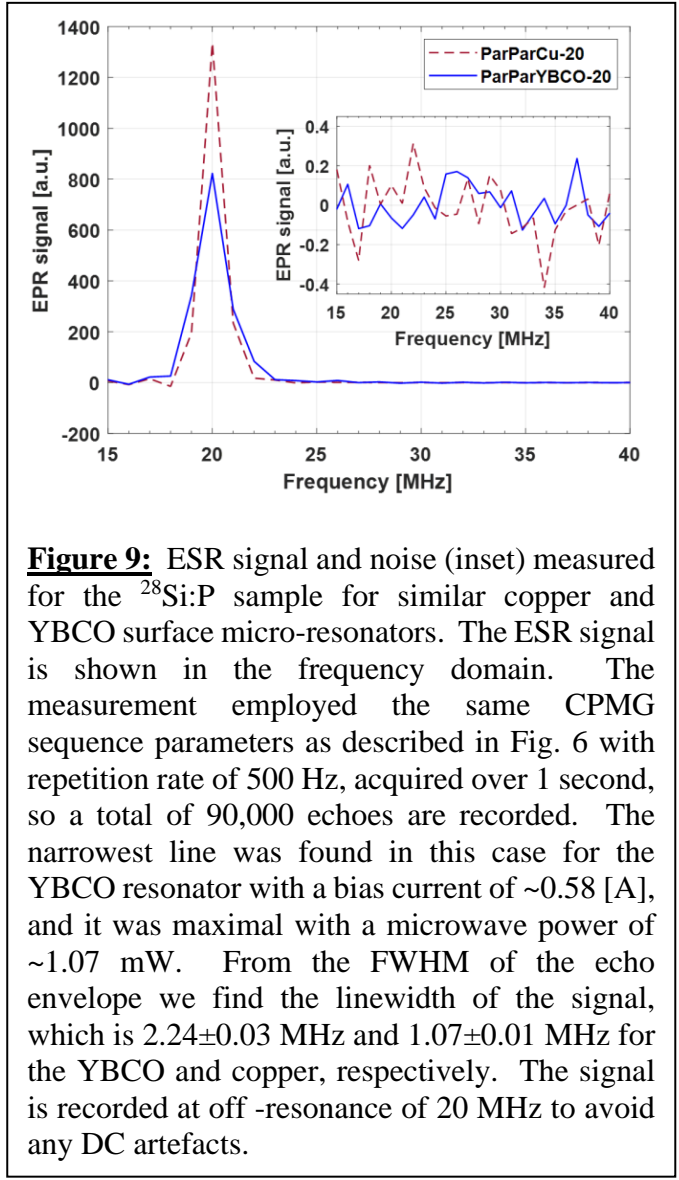

**Figure 9:** ESR signal and noise (inset) measured for the $^{28}$Si:P sample for similar copper and YBCO surface micro-resonators. The ESR signal is shown in the frequency domain. The measurement employed the same CPMG sequence parameters as described in Fig. 6 with repetition rate of 500 Hz, acquired over 1 second, so a total of 90,000 echoes are recorded. The narrowest line was found in this case for the YBCO resonator with a bias current of ~0.58 [A], and it was maximal with a microwave power of ~1.07 mW. From the FWHM of the echo envelope we find the linewidth of the signal, which is 2.24±0.03 MHz and 1.07±0.01 MHz for the YBCO and copper, respectively. The signal is recorded at off-resonance of 20 MHz to avoid any DC artefacts.

## VI.    Discussion and conclusions

This work presents a methodology to design, fabricate, and test YBCO-based superconducting surface micro-resonators. Such resonators can potentially contribute to the sensitive ESR measurement of many types of samples under modest cryogenic conditions as used in mainstream ESR. The Q value we obtained (up to 1670) is more than one order of magnitude better than the one measured for a similar device made of copper. While this value seems to be relatively low compared to previous works with superconductors for ESR, it should be analyzed in view of the relevant context of our motivation. Thus, as noted in the Introduction, some experiments with ESR using miniature superconducting resonators made of Nb achieved similar Q values, but they were limited to very low temperatures and/or very low magnetic fields and thus are irrelevant for most mainstream ESR applications. This is especially true for experiments with aluminum resonators that obtain Q values of more than $10^5$, but require millikelvin temperatures and very low magnetic fields. An additional issue to consider is the fact that we work at relatively high frequencies (~3-7 times higher than other superconducting designs for ESR appearing on the literature), which may also cause increased losses. It is thus clear that although our Q value seems to be low compared to that of other surface superconducting micro-resonators, it is actually very reasonable given the fact that our resonators can operate at much higher temperatures and much higher static field than previous designs. Furthermore, having a Q that is too large would limit the available bandwidth of pulse excitation, which again would not be useful for most mainstream pulsed ESR applications. For example, with the highest Q value we achieved, at ~35 GHz, one can expect to excite a bandwidth of only ~21 MHz. It is possible, of course, to over-couple the resonator, enable a larger bandwidth, and still gain in sensitivity from the high Q values, but the net gain would become negligible as the difference between the over-coupled Q and the unloaded Q becomes very large [39].

Our results also clearly show significant effects caused by a static magnetic field of about 1 T. Such data, which is reported for YBCO resonators for the first time, include decreasing the resonator's quality factor, reducing the ESR signal, broadening its spectrum, and reducing the coherence time of a narrow-line sample placed on the resonator. The latter type of data (reduction in $T_2$) is probably the first-ever investigation of this interesting aspect for superconducting resonators in ESR. The measurement of $T_2$ also opens up the possibility of measuring the current vortices' frequency spectrum (which is an important research field [40]) by, for example, examining $T_2$ decay near the superconductor, as a function of the CPMG interpulse delays, as was done using diamond NV centers in AC magnetometry [41]. We also present a method to mitigate almost all of the magnetic field-induced troubling effects that involves nulling the out-of-plane static magnetic field with an external coil using real-time feedback provided by the ESR signal. This method is easy to implement in standard ESR setups (as opposed to more complex approaches, such as using 3D vector magnets [42]). Still, our efforts could not eliminate these effects completely, resulting in a bottom-line slight loss of SNR when comparing copper and YBCO resonators head-to-head. Thus, while ideally one would expect SNR to improve by a factor of ~2.8 (= $\sqrt{Q_{YBCO}}/\sqrt{Q_{copper}}$), in practice we saw a slight loss in SNR by a factor of ~1.3. This happened even though the noise level for the YBCO resonators was found to be somewhat smaller by a factor of ~1.2. The reasons for the loss in SNR are both the slight increase in line width and the reduction in $T_2$, which caused our CPMG signal (which is preceded by a long 100-μs echo delay) to be smaller and extend to shorter times. Nevertheless, it is probable that for more common samples, which have a much broader natural ESR line and much shorter natural relaxation times, these effects would not be significant and in the bottom line, the YBCO resonators would prevail. One aspect in which we already saw a clear improvement in performance is the significant increase in the resonator's microwave-to-magnetic-field conversion factor, which enabled us to work at power levels that are a factor of ~3.5 smaller than those we needed for the copper resonator. This is in good agreement with the theoretical dependence between this parameter and $\sqrt{Q}$ [3]. The power conversion factor in this case is $C_P$ ~27 G/$\sqrt{W}$ for a sample volume of ~0.17 nL, which also is in good agreement with the theoretical power conversion factor [3]. It should be noted that when using the ParParYBCO-100 resonator, which has a higher Q value and better coupling, we obtained a $C_P$ of ~ 37 G/$\sqrt{W}$ for a sample volume of ~1 nL. This is the highest value reported to-date for such a relatively large sample volume, and it could be very beneficial for mainstream pulsed ESR applications. For example, with 1 W of input microwave power, one can obtain π pulses as short as 5 ns. In Double Electron Electron Resonance (DEER) measurements, a high $B_1$ is very important to obtain a large modulation depth [43]. Large $B_1$ is also important to enable fast spin control [44].

Note about absolute spin sensitivity: In previous work carried out with a ParParCopper-50 resonator, we demonstrated spin sensitivity of ~6.3 × $10^5$ spins/$\sqrt{Hz}$ [23], which is slightly better than the one given here, for a smaller resonator (ParParCopper-20). The reason for this is that in the 2018 experiment we used a very small sample placed only on the bridge, and thus obtained a signal only from the most sensitive region of the resonator, while here we measured a larger volume and thus collected signals also from less sensitive regions. Moreover, the 2018 sample had longer $T_2$ values (since the spin concentration was only $10^{16}$ spin/cm$^3$) which also helped improve its SNR.

Two additional important aspects of our work are the use of an integral miniature cryogenic Hall sensor inside the cryostat to report the off-axis magnetic field close to the sample, and the variable coupling mechanism. The first proved to be very helpful in assessing the level of alignment of our probe head and aided us in the mitigation of the vortices. The second aspect is new with respect to the use of superconducting surface resonators, which often have fixed coupling mechanism. To the best of our knowledge, enabling variable coupling mechanism at cryogenic temperature that can control the coupling rate has not been demonstrated in previous superconducting ESR resonator designs. Variable coupling is also important for the placement of different types of samples on the resonator and to accommodate their different dielectric properties with optimized precision.

It can be concluded that YBCO resonators are essentially "good" in terms of their high Q value and large conversion factor. However, they are "bad" due to the vortices produced on their surface that affect ESR signals. Indeed, this aspect requires additional optimization so that superconductors such as YBCO could be used as a basis for micro-resonator design in mainstream pulsed ESR. Luckily, there seem to be ways to mitigate these issues without too much trouble, so that general-purpose pulsed ESR measurements could potentially be performed with significant net gain at the bottom line. For example, in our current procedure, we cancel the out-of-plane static field component in a specific static field but do not have an automatic mechanism to cancel it in other fields. In the future, such automatic proportional cancellation procedures may be used toto further improve the performance of the device. Other methods that could be employed in the future include incorporating additional means to mitigate static field effects in the resonator's design, such as the use of small strips and patterned surfaces [45-49], or devising resonators with much higher Q values that are placed in a completely closed shield (rather than our semi-open cylindrical shield structure), which may improve upon these results even further. An additional point to consider is that while for most general-purpose ESR applications, resonators with bridge dimensions smaller than ~20 μm do not provide good enough spin concentration sensitivity, for more specialized applications, looking into single spin detection, smaller bridges are advantageous. In such cases, bridges of ~1 μm made of YBCO may not only significantly improve the absolute spin sensitivity but also help mitigate the vortices' effects, since the latter tend to avoid entering into very narrow structures [50-52].

## VII.     Acknowledgments

This work was partially supported by Grant No. 308/17 from the Israel Science Foundation (ISF), Grant No. 1352-302.5 from the German-Israeli Foundation (GIF), Grant No. AZ 98010 from Forschungskooperation Niedersachsen-Israel, and by a grant from the Technion-Waterloo joint program. A.B. acknowledges the Russell Berrie Nanotechnology Institute (RBNI) at the Technion for supporting the clean room activities. The resonators were fabricated at the Technion's Micro-Nano Fabrication & Printing Unit (MNF&PU).